\documentclass[10pt]{article}

\usepackage{amsmath}
\usepackage{amssymb}
\usepackage{graphicx}
\usepackage{dcolumn}
\usepackage{bm}
\usepackage{subfig}
\usepackage{caption}
\usepackage{multirow}
\usepackage{cite}

\usepackage{color}

\makeatletter
\renewcommand{\@biblabel}[1]{\quad#1.}
\makeatother

\date{}

\begin{document}

\begin{flushleft}
{\Large
\textbf{Information filtering in sparse online systems: recommendation via semi-local diffusion}
}
\\
Wei Zeng$^{1,2}$,
An Zeng$^{2,\ast}$,
Ming-Sheng Shang$^{1,3,\ast}$,
Yi-Cheng Zhang$^{1,2,3}$
\\
\bf{1} Web Sciences Center, School of Computer Science and Engineering,
University of Electronic Science and Technology of China, Chengdu 611731, China
\\
\bf{2} Department of Physics, University of Fribourg, Chemin du Mus\'{e}e 3,
Fribourg CH-1700, Switzerland
\\
\bf{3} Institute of Information Economy, Hangzhou Normal University, Hangzhou 310036, China

$\ast$ E-mail: Corresponding an.zeng@unifr.ch, shang.mingsheng@gmail.com
\end{flushleft}

\section*{Abstract}
With the rapid growth of the Internet and overwhelming amount of information and choices that people are confronted with, recommender systems have been developed to effectively support users' decision-making process in the online systems. However, many recommendation algorithms suffer from the data sparsity problem, i.e. the user-object bipartite networks are so sparse that algorithms cannot accurately recommend objects for users. This data sparsity problem makes many well-known recommendation algorithms perform poorly. To solve the problem, we propose a recommendation algorithm based on the semi-local diffusion process on a user-object bipartite network. The numerical simulation on two sparse datasets, Amazon and Bookcross, show that our method significantly outperforms the state-of-the-art methods especially for those small-degree users. Two personalized semi-local diffusion methods are proposed which further improve the recommendation accuracy. Finally, our work indicates that sparse online systems are essentially different from the dense online systems, all the algorithms and conclusions based on dense data should be rechecked again in sparse data.


\section{Introduction}
Owing to the rapid development of the Internet, people are confronted with abundant online contents, which makes it very time-consuming to select the needed information. This is often refereed as the information overload problem. In order to solve it, search engines and recommender systems are wildly investigated and applied to real systems. The search engine returns the relevant contents based on the keywords given by users. Compared to the search engine, the recommender system provides personalized services for users by predicting the potential interests based on their historical choices.

Up to now, many recommendation algorithms have been proposed such as collaborative filtering (CF) \cite{Adomavicius:2005,Shang:EPL:2009,Zeng:2010}, content-based analysis \cite{Pazzani:LNCS:2007} and spectral analysis \cite{Maslov:PRL:2001}. The matrix factorization algorithms have also been wildly investigated by combining high scalability with predictive accuracy \cite{Koren:2009:MF,Hu:08:ICDM}. Recently, some physical processes, including mass diffusion \cite{Zhou:PRE:2007, Zhang:EPL:2007}, heat conduction \cite{Zhang:PRL:2007} and electric circuit analysis \cite{Yang:Plos:2012}, have been applied to design recommendation algorithms. The hybridization of the mass diffusion and heat conduction algorithm is shown to effectively solve the diversity-accuracy dilemma in recommendation \cite{Zhou:PNAS:2010}. Based on these algorithms, many methods have been proposed to further enhance the recommendation diversity and solve the object cold-start problems. For example, the preferential diffusion \cite{Lv:PRE:2011}, the biased heat conduction \cite{Liu:PRE:2011}, network manipulation~\cite{EPL10058005} and the item-oriented method \cite{Qiu:EPL:2011} are shown to be able to largely improve the recommendation accuracy for small-degree objects. More recently, the long-term influence of the hybrid approach on network evolution has been studied~\cite{EPL9718005}.

One of the biggest challenges in recommender systems is the data sparsity problem. That is, the user activity data is too sparse for the recommender system to provide satisfactory recommendations. To solve such sparsity problem, the users' social network is incorporated in the object recommendation. For instance, a random walk model based on both the trust network and user-object bipartite network was designed \cite{Jamali:2009:KDD}. Based on the matrix factorization method, both the user trust network and friendship network can be fused in the object recommendation by regularization \cite{Jamali:2010:RS,Ma:2011:WSDM}. Yang \cite{Yang:2011:WWW} proposed a factor-based random walk model to recommend both online services and friends to users. In addition, the users' membership data (i.e. the social groups that online users joined) is considered and the results indicate that this social information is more valuable than friendship when used to enhance the recommendation accuracy of object \cite{Yuan:2011:RS}.

However, the users' social network is usually much sparser than the user-object network in most systems. More importantly, those users who have collected or purchased few objects might also be inactive in building their social relationships. Therefore, the compensation effect of social networks on the user-object bipartite networks is limited. In this paper, we propose an approach based on the semi-local diffusion process on the user-object bipartite network to solve the data sparsity problem. Our simulation on two real datasets, Amazon and Bookcross, indicates that our method significantly outperforms the state-of-the-art methods especially for these small-degree users. Moreover, two personalized semi-local diffusion methods are proposed which further improve the accuracy. Finally, our work highlights that sparse online systems are essentially different from the dense online systems, all the algorithms and conclusions based on dense data should be reconsidered and reexamined in sparse data.

\section{Data sparsity problem}
\label{Sec:problem}
The hybrid method in ref. \cite{Zhou:PNAS:2010}
takes into account both the mass diffusion \cite{Zhou:PRE:2007} and the heat conduction \cite{Zhang:PRL:2007} process. This method is shown to be able to provide not only accurate but also diverse recommendations for users when applied to dense datasets. Here, we argue that this hybrid method fails in sparse datasets. As an example, we test this hybrid method on two sparse datasets: Amazon and Bookcross. \emph{Amazon.com} is a multinational e-commerce company and the world's largest online retailer. \emph{Bookcrossing.com} is a book sharing web site where book lovers can exchange their books and experiences with each other. Some basic statistics of these two datasets are presented in the Table \ref{Tab:data_statistics}. Each data is randomly divided into two parts: the training set ($E^T$) and the probe set ($E^P$). The training set contains $80\%$ of the original data and the recommendation algorithm runs on it. The rest of the data forms the probe set, which will be used to examine the recommendation performance.

Online commercial systems can be naturally described by the user-item bipartite networks with the adjacency matrix $A$ in which the element $a_{i\alpha}=1$ if the user $i$ has collected the item $\alpha$, and $a_{i\alpha}=0$ otherwise (throughout this paper we use Greek and Latin letters, respectively, for item- and user-related indices)~\cite{Shang:EPL:2009,Zhou:Plos:2011}. When recommending objects for user $i$, the hybrid method works by assigning each object collected by user $i$ one unit of resource. The initial resources are denoted by the vector $\overrightarrow{f}$ where $f_{\alpha}$ is the resource possessed by object $\alpha$. Then they will be redistributed via the transformation $\overrightarrow{f'}=W\overrightarrow{f}$, where
\begin{equation}
W_{\alpha\beta}=\frac{1}{k_{\alpha}^{1-\lambda}k_{\beta}^{\lambda}}\sum_{j=1}^{N}\frac{a_{j\alpha}a_{j\beta}}{k_j}
\label{Eq:hybrid_w}
\end{equation}
is the redistribution matrix, with $k_{\alpha}=\sum_{l=1}^{N}a_{l\alpha}$ and $k_j=\sum_{\gamma=1}^{M}a_{j\gamma}$ denoting the degree of object $\alpha$ and user $j$, respectively. $N$ and $M$ are the number of users and objects, respectively. $\lambda$ is a tunable parameter which adjusts the relative weight between the mass diffusion algorithm ($\lambda=1$) and heat conduction algorithm ($\lambda=0$). The resulting recommendation list of uncollected items is sorted according to $\overrightarrow{f'}$ in descending order.

In order to measure the recommendation accuracy, we make use of the ranking score ($RS$). Specifically, $RS$ measures whether the ordering of the items in
the recommendation list matches the users' real preference. As discussed above, the recommender system will provide each user with a ranking list which contains all his uncollected items. For a target user $i$, we calculate the ranking position for each of his link in the probe set. Suppose
one of his uncollected items $\alpha$ is ranked at the $5th$ place
and the total number of his uncollected items is $100$, the
ranking score $RS_{i\alpha}$ will be $0.05$. In a good recommendation,
the items in the probe set should be ranked higher, so that
$RS$ will be smaller. Therefore, the mean value of the $RS$ over
all the user-item relations in the probe set can be used to
evaluate the recommendation accuracy as
\begin{equation}
\langle RS\rangle=\frac{1}{|E_{P}|}\sum_{i\alpha\in E_P}RS_{i\alpha}
\end{equation}
The smaller the value of $\langle RS\rangle$, the higher the recommendation
accuracy.

\begin{figure}
\centering
\includegraphics[width=9cm,height=5cm]{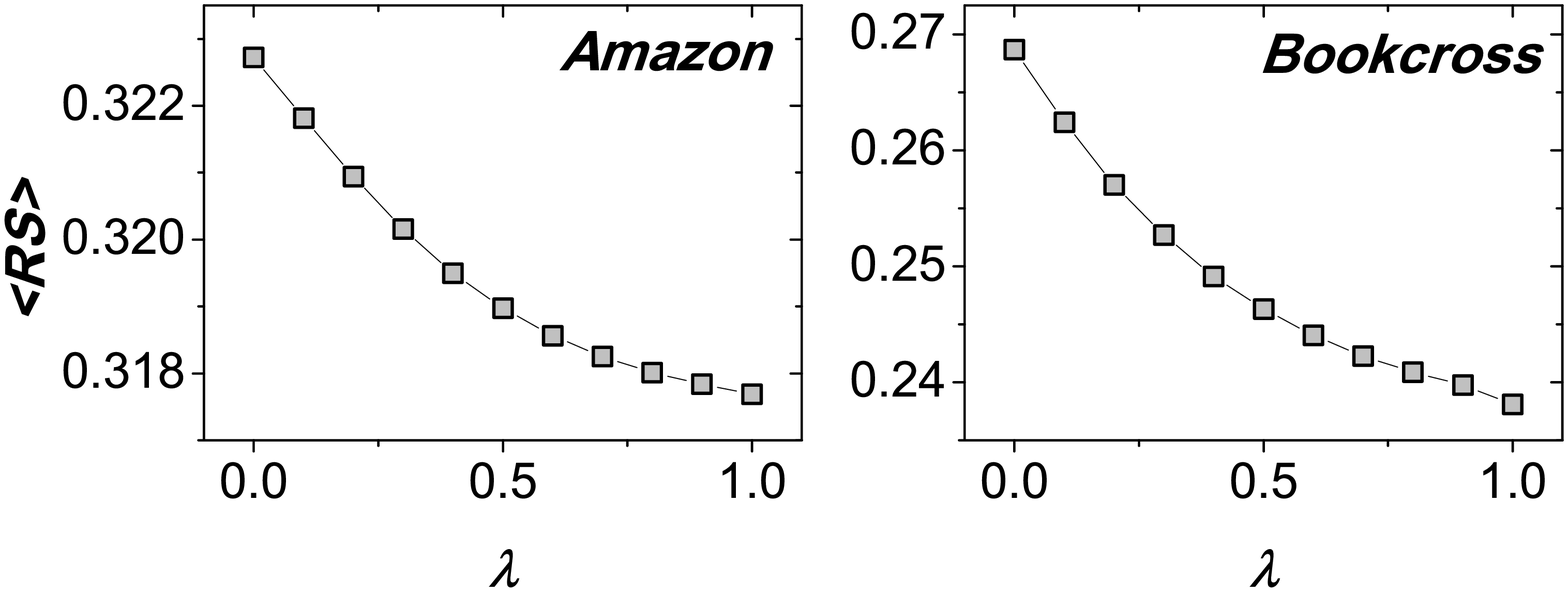}
\caption{The ranking score of the hybrid method on Amazon and Bookcross. $\lambda$ is used to tune the contribution of the heat conduction and the mass diffusion process. When $\lambda=1$, the hybrid method gives the pure mass diffusion method and $\lambda=0$ it degenerates to pure heat conduction method (more details about the hybrid method can be found in \cite{Zhou:PNAS:2010}).}
\label{Fig:hybrid_rs}
\end{figure}

In ref.~\cite{Zhou:PNAS:2010}, $\langle RS\rangle$ can achieve an optimal value when adjusting the parameter $\lambda$ of the hybrid recommendation method. However, when applied to the sparse data mentioned above, $\langle RS\rangle$ changes monotonously with $\lambda$, as presented in fig.~\ref{Fig:hybrid_rs}. In other words, the recommendation accuracy cannot be improved by taking into account the heat conduction process in the mass diffusion method.

\begin{figure}[!ht]
\centering
\includegraphics[width=9cm,height=5cm]{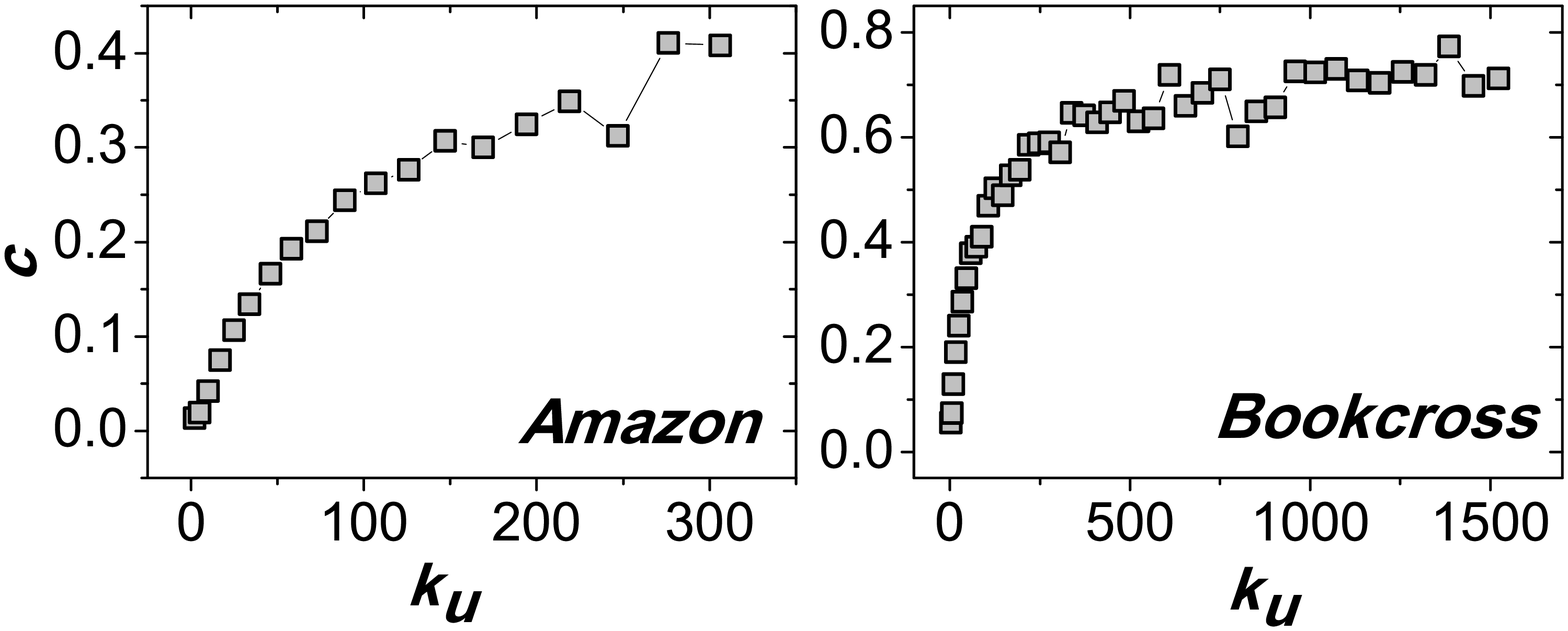}
\caption{Dependance of the converge $c$ on the user degree in Amazon and Bookcross. For a given $x$, its corresponding $c$ is obtained by averaging all the users whose degrees are in the range of $[a(x^2-x), a(x^2+2)]$, where $a$ is chosen as $\frac{1}{2}\log5$.}
\label{Fig:converge}
\end{figure}

To understand the reason, we introduce a concept called \emph{coverage}, $c$. Given a random walker starting from $u$, we denote $\pi^{(t)}_{\alpha}$ as the probability that the walker reaches item $\alpha$ after $t$ step. Then the \emph{coverage} $c_u$ is defined as the ratio of the number of objects whose $\pi^{(3)}_{\alpha}$ is larger than 0. If $\pi^{(3)}_{\alpha}=0$, the resource item $\alpha$ received in these 3-step diffusion-based algorithms (e.g. diffusion process and heat conduction) will equal to $0$. The average coverage $\overline{c}$ over all users are $0.0301$ for Amazon and $0.1413$ for Bookcross, respectively. In other words, the resources most objects received will equal to $0$ if we choose the 3-step diffusion-based recommendation algorithms. Note that the hybridization \cite{Zhou:PNAS:2010} only changes the amount of resource of the items whose $\pi^{(3)}_{\alpha}$ are lager than $0$. The resource of the items whose $\pi^{(3)}_{\alpha}=0$ will stay $0$ under all hybrid parameter. Since the coverage dominates the recommendation accuracy in sparse data, the hybrid method cannot improve the recommendation accuracy as shown in fig.~\ref{Fig:hybrid_rs}. Moreover, we show the relationship between the user degree and the coverage $c$ in the fig. \ref{Fig:converge}. The coverage nonlinearly decreases with user degree, which leads to a even more serious user cold-start problem. In next section, we will propose a semi-local diffusion method to effectively increase the diffusion coverage and break the tie among these items with $0$ resource.

\section{Algorithm and metrics}
Our semi-local diffusion method will be directly built on the mass diffusion method (short for MD method)\cite{Zhou:PRE:2007}. The MD method is simply the case when $\lambda=1$ in the hybrid method. Given a target user $u$, the first step of MD is to allocate one unit resource to each of $u$'s collected items. After 2 more steps of diffusion, the resource will be back to the item side. For convenience, we denote every $2$ steps after the $1$-step as one macro-step ($MS$ for short) of diffusion. The original $3$-step diffusion is combined by the first step (the initial resources allocating process) and $1$ macro-step diffusion. As discussed above, the original $3$-step diffusion method suffers from the data sparsity problem since most objects' resources are $0$. To solve this problem, we let the resources diffuse on the bipartite network more than one macro-step. Given the target user $u$, $\overrightarrow{f}$ is the initial resource vector. After one macro-step, items' resource can be expressed as $\overrightarrow{f}^{(1)} = W\overrightarrow{f}$, where $W$ is the resource redistribution matrix for mass diffusion algorithm (with $\lambda=1$ in equation \ref{Eq:hybrid_w}). Likewise, we can calculate items' resource after $n$ macro-steps of diffusion as $\overrightarrow{f}^{(n)}=W\overrightarrow{f}^{(n-1)}=W^n\overrightarrow{f}$. To recommend objects to $u$, one can sort the $\overrightarrow{f}^{(n)}$ in descending order and those objects with most resources will be recommended. Since the algorithm above uses less than global information but a bit more than pure local information, we call this method as Semi-Local Diffusion (SLD) recommendation method.

In previous section, we used the ranking score to measure the recommendation accuracy. Since real users usually consider only the top part of the recommendation list, a more practical measure should take into account the number of a user's hidden links contained in the top-L places. Therefore, we use another recommendation accuracy measure called ``recall". For a user $i$, the number of his/her probe set links appearing in the top-$L$ recommendation list is denoted as $d_i(L)$, these links are also called hit links of user $i$. The total number of his/her probe set links is denoted as $N_i$. The recall of user $i$ is defined as
\begin{equation}
\label{eq:recall_def}
Re_i(L) = d_i(L)/N_i.
\end{equation}
The recall of the whole system is defined as
\begin{equation}
Re(L) = \frac{1}{N}\sum_{i=1}^N{Re_i(L)}.
\end{equation}
A higher recall value indicates a higher accuracy of recommendation.

\section{Result Analysis}
\label{Sec:result}
\begin{figure}[!ht]
\centering
\includegraphics[width=9cm,height=5cm]{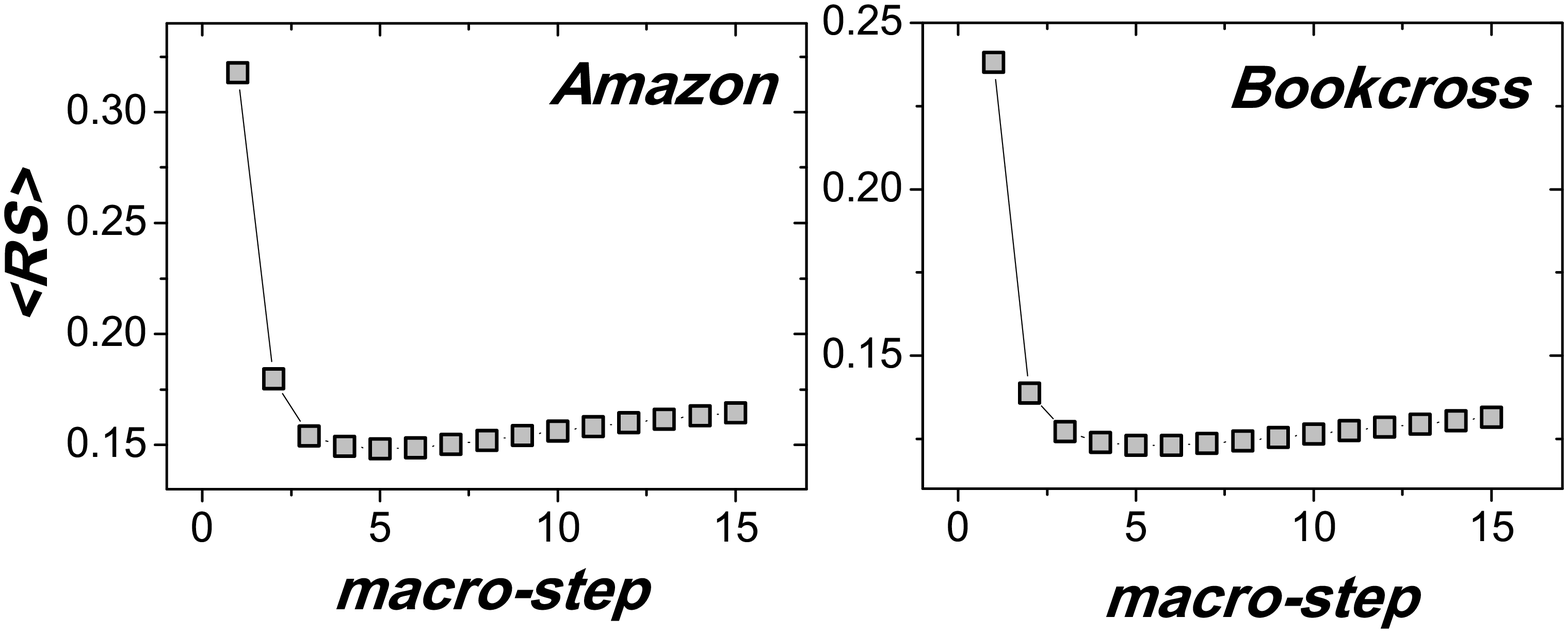}
\caption{The ranking score $\langle RS\rangle$ of the semi-local diffusion method in Amazon and Bookcross. For both datasets, we obtain the lowest ranking score when the macro diffusion step is 5.}
\label{Fig:msd_rs}
\end{figure}

\begin{figure}[!ht]
\centering
\includegraphics[width=9cm,height=7cm]{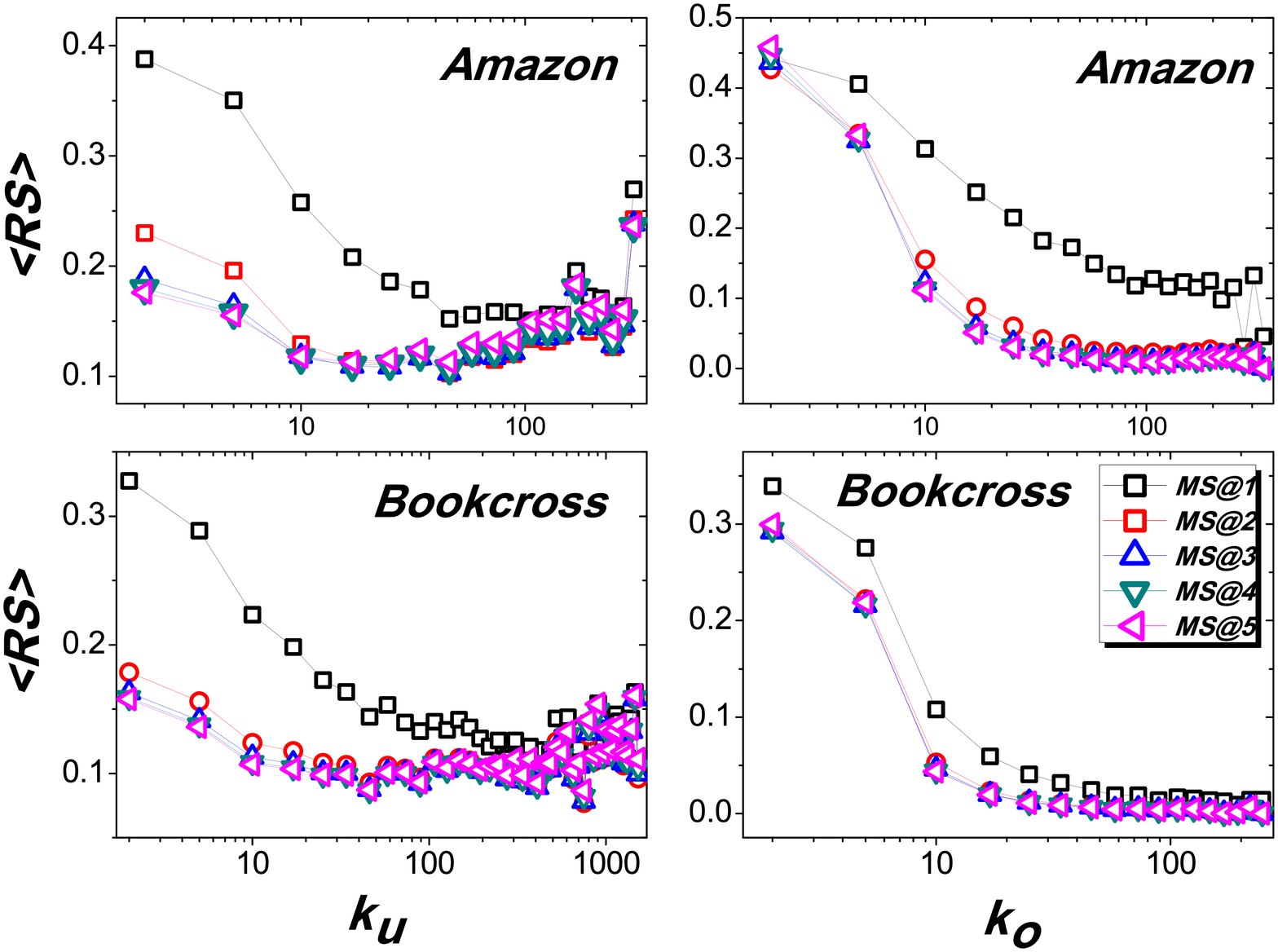}
\caption{Dependence of the ranking score $\langle RS\rangle$ on user degree and object degree. The $MS@T$ means the macro-step of the diffusion is $T$. For a given $x$, its corresponding $\langle RS\rangle$ is obtained by averaging all the users (or objects) whose degrees are in the range of $[a(x^2-x), a(x^2+2)]$, where $a$ is chosen as $\frac{1}{2}\log5$.}
\label{Fig:msd_degree_rs}
\end{figure}

If we let the objects' resources diffuse on the bipartite network for multiple macro-steps, more objects in the probe set will have final resources larger than 0, which results in higher rank of these objects and thus a lower overall ranking score $\langle RS\rangle$. The relation between the overall $\langle RS\rangle$ and the number of macro-steps is presented in fig.~\ref{Fig:msd_rs}. If \emph{macro-step=1}, the method degenerates to the standard Mass diffusion method. From the figure, one can see that the ranking score is improved significantly by the the SLD method and the optimal diffusion macro-step is $5$ in both datasets. If the macro-step is more than $5$, the ranking score gets worse but still much better than the original Mass diffusion method. Additionally, we report the dependence of the ranking score on the user degree and object degree in fig.~\ref{Fig:msd_degree_rs}. The left two figures of fig.~\ref{Fig:msd_degree_rs} give the relationship between the user degree and the ranking score. One can see that the ranking scores of small-degree users who have collected few objects are improved greatly since these users' \emph{coverage} of objects are increased significantly by the SLD. The right two figures of fig. \ref{Fig:msd_degree_rs} show the relationship between the object degree and the ranking score. It can be seen that the SLD can improve the ranking scores of both the small-degree and large-degree objects.

\begin{figure}[!ht]
\centering
\includegraphics[width=9cm,height=5cm]{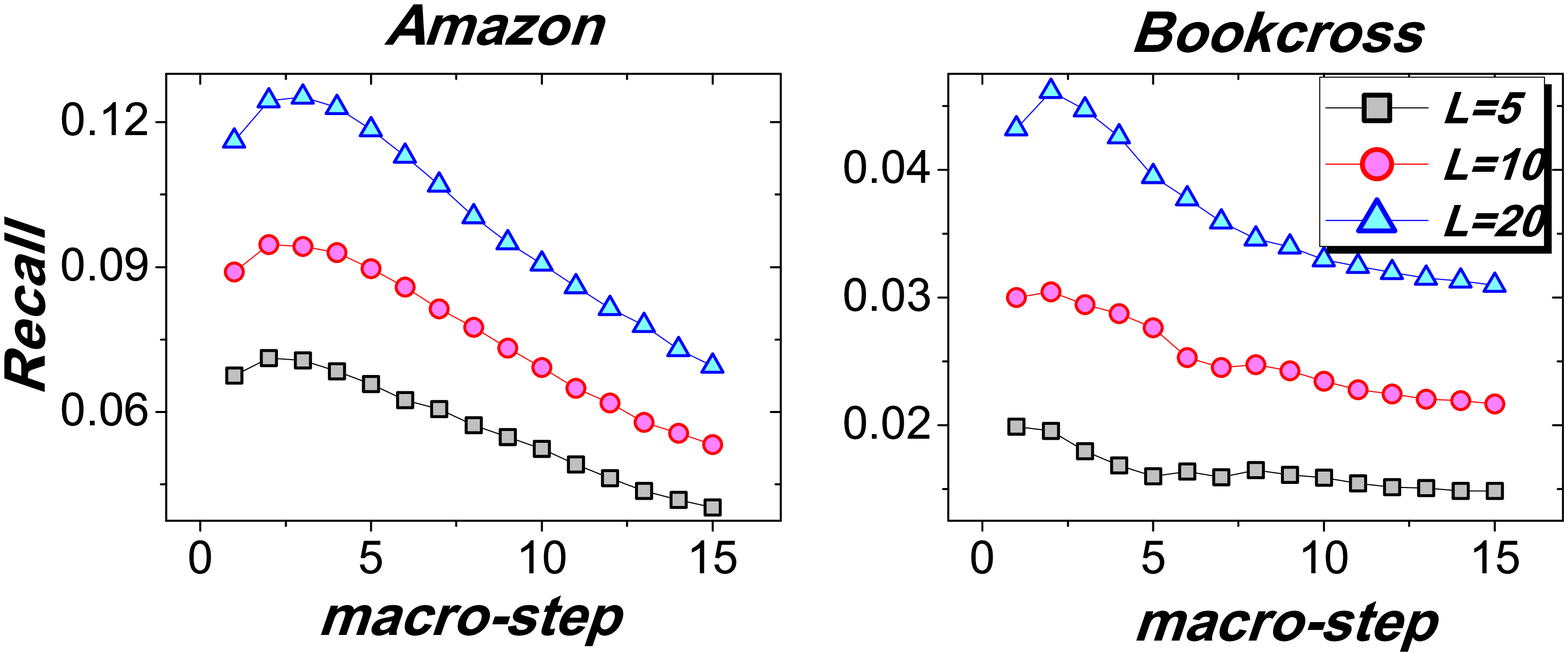}
\caption{The recall of the semi-local diffusion method in Amazon and Bookcross. For both datasets, we obtain the best performance when the macro-step of the diffusion is 2.}
\label{Fig:msd_accuracy}
\end{figure}

\begin{figure}[!ht]
\centering
\includegraphics[width=9cm,height=7cm]{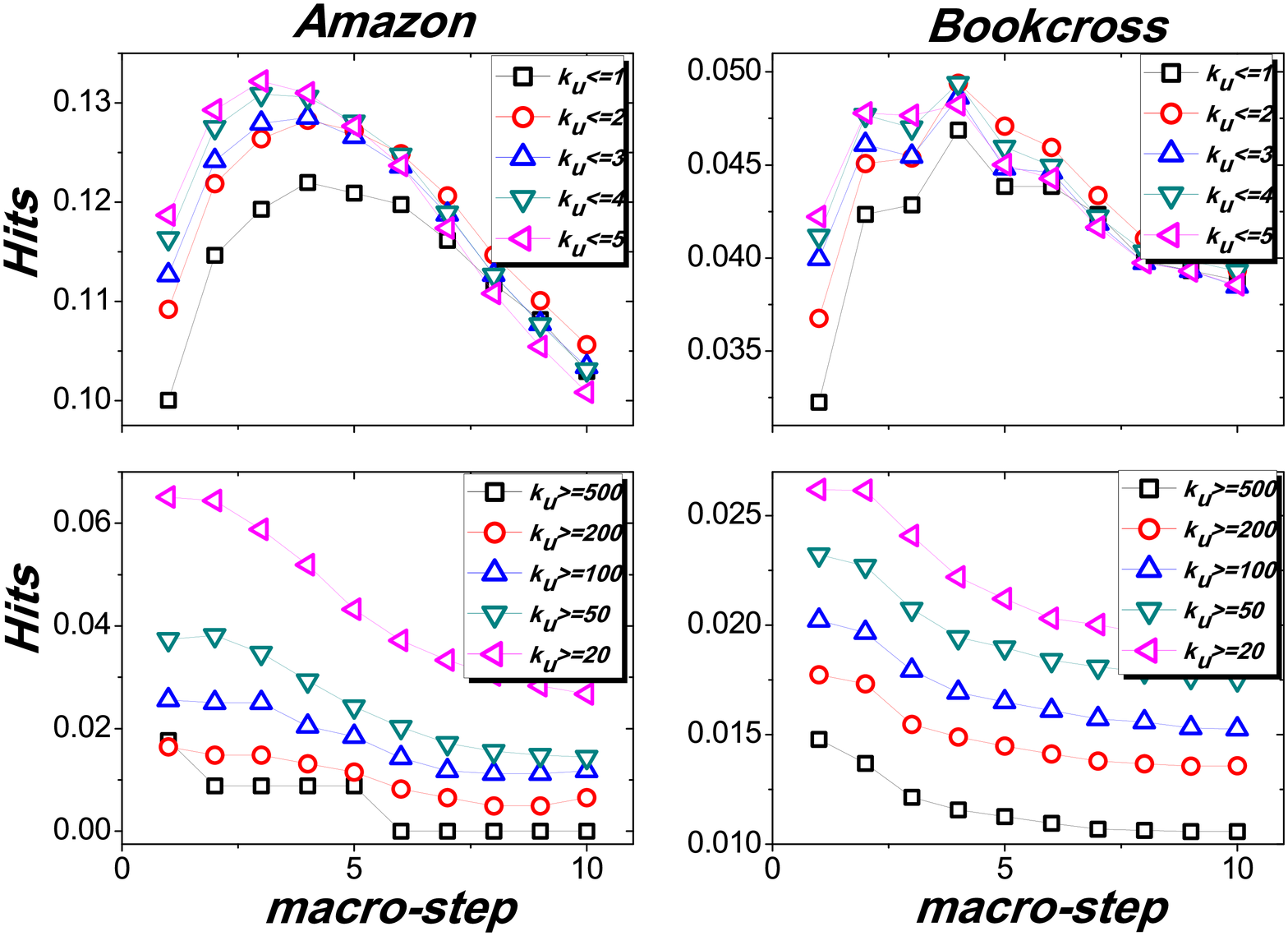}
\caption{Dependence of \emph{Hits} on the diffusion macro-step. The recommendation list length $L$ is set to 20. $k_u<=D$ means that we only consider the users whose degree is no larger than $D$.}
\label{Fig:msd_user_hits}
\end{figure}

\begin{figure}[!ht]
\centering
\includegraphics[width=9cm,height=5cm]{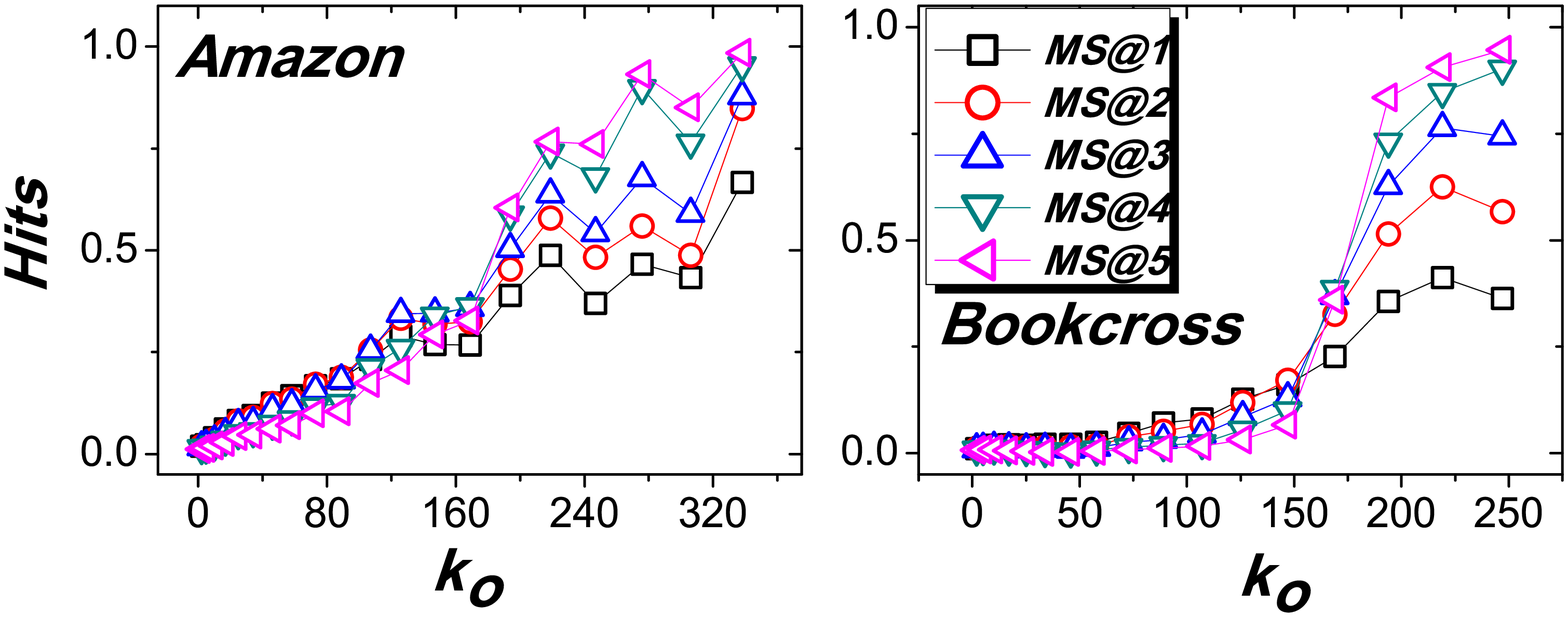}
\caption{The relationship between object degree and \emph{Hits}. The recommendation list length $L$ is set to 20. $MS@T$ means the macro-diffusion step is $T$.}
\label{Fig:msd_itemdegree_hits}
\end{figure}

Another interesting question is whether the accuracy of top-$L$ recommendation list will be improved the same as the ranking score by the SLD. The relation between the recall and the number of macro-steps is presented in fig. \ref{Fig:msd_accuracy}. As one can see, for both datasets, we get the best performance when the macro-step is $2$. However, when the macro-step is more than $2$, the recall of both datasets start to decrease. To uncover the reason, we study in detail the relationship between the top-$L$ accuracy and user degree and object degree, respectively. Since recall is usually used in measuring the overall top-L accuracy, we choose $Hits$ to measure the top-L accuracy for a set of users or items. For a set of nodes (users/items), $Hits$ is defined as the ratio of their hit links $D(L)$ to the total number of links in the probe set for these nodes $|E_p|$. Mathematically, it reads
\begin{equation}
\label{Eq:hits}
H(L) = D(L) / |E_p|,
\end{equation}
Like the $recall$, a higher $Hits$ indicates a better recommendation. The top two subfigures of fig. \ref{Fig:msd_user_hits} show the $Hits$ of the users whose degrees are no larger than $5$. It can be seen that the accuracy of top-$20$ recommendation lists of those inactive users are improved considerably by the SLD if the macro-step of diffusion is less than $5$. The best macro-step is $3$ for Amazon and $4$ for Bookcross, respectively. If the macro-step of diffusion is larger than $5$, the $Hits$ of those users start to decrease. The bottom two subfigures of fig. \ref{Fig:msd_user_hits} give the $Hits$ of the users whose degrees are no smaller than 20. It shows that the $Hits$ decreases monotonously with macro-step. In addition, we plot the relationship between $Hits$ and the object degree in fig. \ref{Fig:msd_itemdegree_hits}. It shows that the SLD method tends to improve the $Hits$ of large-degree objects. Generally speaking, small degree users incline to select popular items~\cite{Shang:EPL:2010}. However, since the small degree users only have limit number of links, the original $3$-step diffusion cannot reach the relevant popular items for them. On the other hand, the SLD method effectively increases the diffusion coverage and discover the most relevant popular items for these small degree users. This is of great importance from practical point of view since these new/inactive users are very sensitive to the quality of recommendation and poor recommendations may lead to losing them from the website.

Our result above shows that the high order diffusion resources may play different role in the recommendation for users and objects with different degrees. Therefore, the information of the high order diffusion should be used in a personalized way. Accordingly, we propose two extended recommendation methods: the user-based semi-local diffusion method (U-SLD for short) and the object-based semi-local diffusion method (O-SLD for short). We denote $\overrightarrow{f}^{(1)}, \overrightarrow{f}^{(2)}, ..., \overrightarrow{f}^{(n)}$ as the final resource vectors after \emph{1, 2, ..., n} macro-steps of diffusion, respectively. $\overrightarrow{f}^{(n)}$ can be easily calculated by $\overrightarrow{f}^{(n)}=W\overrightarrow{f}^{(n-1)}=W^n\overrightarrow{f}$. Given the target user $u$, the user-based semi-local diffusion method is to combine these $n$ resource vectors based on $u$'s degree. Mathematically, the final score of object $\alpha$ reads
\begin{equation}
\label{Eq:u_msd}
F_{\alpha}^{u} = f^{(1)}_{\alpha}+\sum_{i=2}^n{\frac{1}{k_u^{\theta}}f_{\alpha}^{(i)}},
\end{equation}
where $k_u$ is $u$'s degree and $\theta$ is a free parameter to tune the weight of $\overrightarrow{f}^{(i)}$ ($i\geq 2$) based on $u$'s degree. If $\theta>0$, the second term will play a more significant role when recommending objects for small-degree users, and vice versa.

\begin{figure}[!ht]
\centering
\includegraphics[width=9cm,height=7cm]{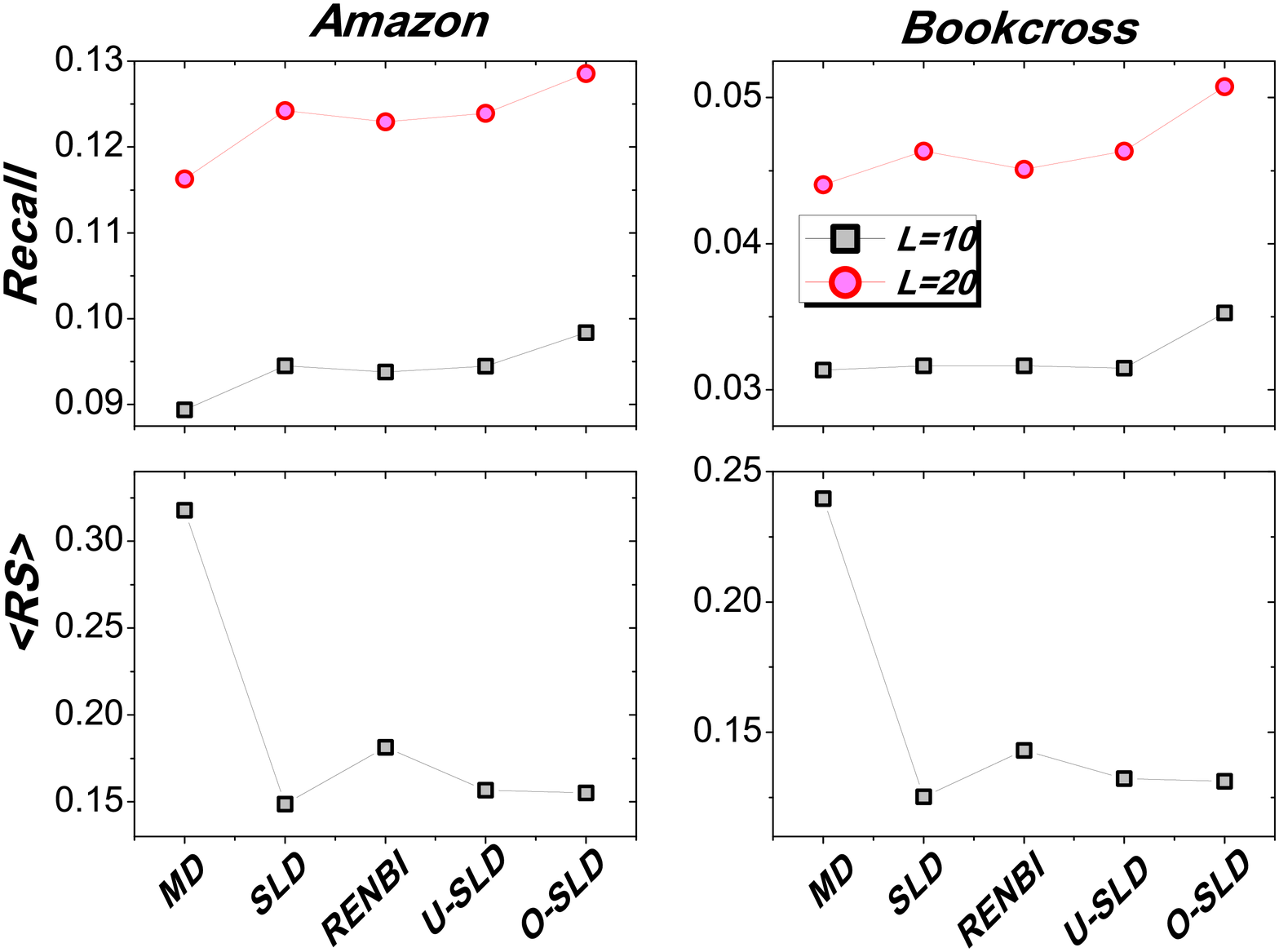}
\caption{The accuracy comparison of different algorithms.}
\label{Fig:method_compare}
\end{figure}

In the sparse dataset, the coverage of $3$-step diffusion is very low. Even some popular items cannot be effectively reached by users. The object-based semi-local diffusion method accumulates those resources based on the object degree. The final score of object $\alpha$ computed by this method is
\begin{equation}
\label{Eq:o_msd}
F_{\alpha}^{u}=f^{(1)}_{\alpha}+\sum_{i=2}^n{\frac{1}{k_{\alpha}^{\theta}}f_{\alpha}^{(i)}}.
\end{equation}
If $\theta>0$, the second term will play a more significant role in calculating the score for small-degree items, and vice versa. We sort the vector $F^{u}$ in descending order and those objects with highest scores will be recommended to $u$. The results on Amazon and Bookcross are reported in fig. \ref{Fig:method_compare} and the optimal parameters $\theta$ of algorithms discussed above are presented in table \ref{Tab:optimal_parameter}. In order to balance the improvement on $ranking score$ and $recall$, we set $n=3$ in both U-SLD and O-SLD.

Actually, similar idea has been applied to eliminate the redundant correlations in dense datasets~\cite{Zhou:NJP:2009}. The method in~\cite{Zhou:NJP:2009} is called RENBI method and defined as
\begin{equation}
\label{Eq:RENBI_method}
\overrightarrow{f'}=(W+\theta W^2)\overrightarrow{f},
\end{equation}
where the elements of matrix $W$ are defined by eq. \ref{Eq:hybrid_w} with $\lambda=1$, $\overrightarrow{f'}$ and $\overrightarrow{f}$ is the final resource vector and the initial resource vector, respectively, and $\theta$ is a free parameter. In~\cite{Zhou:NJP:2009}, the authors focus on improving the accuracy and diversity of recommendation by eliminating the redundant information and they find that the optimal $\theta$ defined in eq. \ref{Eq:RENBI_method} is negative. However, the information of high order diffusion is not redundant any more in sparse dataset. Moreover, the RENBI method is not personalized since the weight of high order diffusion resources is the same for all users. We will compare the U-SLD and O-SLD methods to the RENBI method.

The top subfigures of fig.~\ref{Fig:method_compare} show the results of \emph{recall} in Amazon and Bookcross. Clearly, the recall of SLD is much higher than that of MD in both datasets. This is because the recommendation accuracy of small-degree users is significantly improved by SLD. Moreover, the RENBI method is also better than the MD method, but it is worse than the SLD. From the table \ref{Tab:optimal_parameter}, we can also see that the optimal $\theta$ in Eq. \ref{Eq:RENBI_method} are $0.9$ for Amazon and $0.7$ for Bookcross, respectively. This is different from the result in ref \cite{Zhou:NJP:2009} where the method is tested in dense data and the optimal $\theta$ is found to be negative. Our results indicate that the information of high order diffusion is in fact not redundant information in the sparse data. Both the U-SLD and O-SLD methods are better than the RENBI method in recall. The improvement is due to the personalized use of the high order diffusion information. Finally, we can see that the O-SLD achieves the best \emph{recall} among these methods and the optimal $\theta$ defined in Eq. \ref{Eq:o_msd} is negative in both datasets from table \ref{Tab:optimal_parameter}. That is to say, the information of high order diffusion should be considered more on the large degree items than small degree items. This is because small degree users inclines to select the popular items while these items cannot be effectively reached by one macro-step diffusion. Note that once those small degree users have selected many objects, we could then recommend diverse objects to them.

The bottom subfigures of fig. \ref{Fig:method_compare} show the results of \emph{ranking score} in Amazon and Bookcross. One can see that the ranking score of SLD method is much lower than that of MD. From the table \ref{Tab:optimal_parameter}, it is shown that the optimal diffusion step is $5$ in both datasets. RENBI also achieves a considerable improvement in ranking score compared to MD, but its ranking score is higher than that of SLD. The optimal $\theta$ of RENBI is also positive in both datasets. This supports again that the high order diffusion information is actually useful in enhancing the recommendation accuracy in sparse data. Although the ranking score of U-SLD and O-SLD method are slightly higher than the SLD method, these two methods enjoy a much better ranking score than RENBI. Taken together the results of ranking score and recall, O-SLD seems to be the best recommendation algorithm in sparse data. It provides not only a good ranking of users' unselected objects but also an accurate top-$L$ recommendation list.

\section{Discussion}
\label{conclusion}
The data sparsity problem is one of the biggest challenges in recommender systems. There are a large number of online users and objects with very few connections, which leads to the poor performance of many well-known recommendation algorithms. However, the data sparsity problem has not yet been systematically studied and not yet well addressed. Take the hybrid method \cite{Zhou:PNAS:2010} for example, one cannot get an improved recommendation accuracy when combining the mass diffusion and heat conduction algorithms. As a matter of fact, the data of most real online systems is much sparser than the data used in this paper. Therefore, solving the data sparsity problem is of great significance from the practical point of view.

In this paper, we propose a semi-local diffusion (SLD) method to solve the data sparsity problem in recommender systems. The results on two real online datasets indicate that our method significantly outperforms other well-known algorithms. Two personalized semi-local diffusion methods are also proposed which further improve the accuracy. Our analysis shows that the recommendation accuracy of small-degree users is greatly improved by the SLD method. In practical use, it can largely improve the user experience of the new comers, so that more users will be attracted by the web site. Finally, we remark that sparse online system are essentially different from the dense online system, all the algorithms and conclusions based on dense data should be rechecked in sparse data.

\section*{Acknowledgments}
This work is supported by the opening foundation of Institute of Information Economy in Hangzhou Normal University (Grant No. PD12001003002002) and the Sichuan Provincial Science and Technology Department (Grant No. 2012FZ0120). W.Z. acknowledges the support from Sino-Swiss Science and Technology Cooperation Program (EG57-092011). A.Z. acknowledges the support from China scholarship Council.

\bibliography{refs}

\begin{thebibliography}{10}
\providecommand{\url}[1]{\texttt{#1}}
\providecommand{\urlprefix}{URL }
\expandafter\ifx\csname urlstyle\endcsname\relax
  \providecommand{\doi}[1]{doi:\discretionary{}{}{}#1}\else
  \providecommand{\doi}{doi:\discretionary{}{}{}\begingroup
  \urlstyle{rm}\Url}\fi
\providecommand{\bibAnnoteFile}[1]{%
  \IfFileExists{#1}{\begin{quotation}\noindent\textsc{Key:} #1\\
  \textsc{Annotation:}\ \input{#1}\end{quotation}}{}}
\providecommand{\bibAnnote}[2]{%
  \begin{quotation}\noindent\textsc{Key:} #1\\
  \textsc{Annotation:}\ #2\end{quotation}}
\providecommand{\eprint}[2][]{\url{#2}}

\bibitem{Adomavicius:2005}
Adomavicius G, Tuzhilin A (2005) Toward the next generation of recommender
  systems: A survey of the state-of-the-art and possible extensions.
\newblock IEEE Trans Knowl Data Eng 17: 734--749.
\bibAnnoteFile{Adomavicius:2005}

\bibitem{Shang:EPL:2009}
Shang MS, L\"u LY, Zeng W, Zhang YC, Zhou T (2009) Relevance is more
  significant than correlation: Information filtering on sparse data.
\newblock EPL 88: 68008.
\bibAnnoteFile{Shang:EPL:2009}

\bibitem{Zeng:2010}
Zeng W, Shang MS, Zhang QM, L\"u LY, Zhou T (2010) Can dissimilar users
  contribute to accuracy and diversity of personalized recommendation?
\newblock Int J Mod Phys C 21: 1217-1227.
\bibAnnoteFile{Zeng:2010}

\bibitem{Pazzani:LNCS:2007}
Pazzani MJ, Billsus D (2007) The adaptive web.
\newblock Berlin, Heidelberg: Springer-Verlag, chapter Content-based
  recommendation systems. pp. 325--341.
\bibAnnoteFile{Pazzani:LNCS:2007}

\bibitem{Maslov:PRL:2001}
Maslov S, Zhang YC (2001) Extracting hidden information from knowledge
  networks.
\newblock Phys Rev Lett 87: 248701.
\bibAnnoteFile{Maslov:PRL:2001}

\bibitem{Koren:2009:MF}
Koren Y, Bell R, Volinsky C (2009) Matrix factorization techniques for
  recommender systems.
\newblock Computer 42: 30--37.
\bibAnnoteFile{Koren:2009:MF}

\bibitem{Hu:08:ICDM}
Hu YF, Koren Y, Volinsky C (2008) Collaborative filtering for implicit feedback
  datasets.
\newblock In: Proceedings of the 2008 Eighth IEEE International Conference on
  Data Mining. ICDM '08, pp. 263--272.
\bibAnnoteFile{Hu:08:ICDM}

\bibitem{Zhou:PRE:2007}
Zhou T, Ren J, Medo M, Zhang YC (2007) Bipartite network projection and
  personal recommendation.
\newblock Physical Review E 76: 046115.
\bibAnnoteFile{Zhou:PRE:2007}

\bibitem{Zhang:EPL:2007}
Zhang YC, Medo M, Ren J, Zhou T, Li T, et~al. (2007) Recommendation model based
  on opinion diffusion.
\newblock EPL 80: 68003.
\bibAnnoteFile{Zhang:EPL:2007}

\bibitem{Zhang:PRL:2007}
Zhang YC, Blattner M, Yu YK (2007) Heat conduction process on community
  networks as a recommendation model.
\newblock Phys Rev Lett 99: 154301.
\bibAnnoteFile{Zhang:PRL:2007}

\bibitem{Yang:Plos:2012}
Yang J, Kim J, Kim W, Kim YH (2012) Measuring user similarity using electric
  circuit analysis: Application to collaborative filtering.
\newblock PLoS ONE 7: e49126.
\bibAnnoteFile{Yang:Plos:2012}

\bibitem{Zhou:PNAS:2010}
Zhou T, Kuscsik Z, Liu JG, Medo M, Wakeling JR, et~al. (2010) Solving the
  apparent diversity-accuracy dilemma of recommender systems.
\newblock Proc Natl Acad Sci USA 107: 4511-4515.
\bibAnnoteFile{Zhou:PNAS:2010}

\bibitem{Lv:PRE:2011}
L\"u LY, Liu WP (2011) Information filtering via preferential diffusion.
\newblock Phys Rev E 83: 066119.
\bibAnnoteFile{Lv:PRE:2011}

\bibitem{Liu:PRE:2011}
Liu JG, Zhou T, Guo Q (2011) Information filtering via biased heat conduction.
\newblock Phys Rev E 84: 037101.
\bibAnnoteFile{Liu:PRE:2011}

\bibitem{EPL10058005}
Zhang FG, Zeng A (2012) Improving information filtering via network
  manipulation.
\newblock EPL 100: 58005.
\bibAnnoteFile{EPL10058005}

\bibitem{Qiu:EPL:2011}
Qiu T, Chen G, Zhang ZK, Zhou T (2011) An item-oriented recommendation
  algorithm on cold-start problem.
\newblock EPL 95: 58003.
\bibAnnoteFile{Qiu:EPL:2011}

\bibitem{EPL9718005}
Zeng A, Yeung CH, Shang MS, Zhang YC (2012) The reinforcing influence of
  recommendations on global diversification.
\newblock EPL 97: 18005.
\bibAnnoteFile{EPL9718005}

\bibitem{Jamali:2009:KDD}
Jamali M, Ester M (2009) Trustwalker: a random walk model for combining
  trust-based and item-based recommendation.
\newblock In: Proceedings of the 15th ACM SIGKDD international conference on
  Knowledge discovery and data mining. KDD '09, pp. 397--406.
\bibAnnoteFile{Jamali:2009:KDD}

\bibitem{Jamali:2010:RS}
Jamali M, Ester M (2010) A matrix factorization technique with trust
  propagation for recommendation in social networks.
\newblock In: Proceedings of the fourth ACM conference on Recommender systems.
  RecSys '10, pp. 135--142.
\bibAnnoteFile{Jamali:2010:RS}

\bibitem{Ma:2011:WSDM}
Ma H, Zhou DY, Liu C, Lyu MR, King I (2011) Recommender systems with social
  regularization.
\newblock In: Proceedings of the fourth ACM international conference on Web
  search and data mining. WSDM '11, pp. 287--296.
\bibAnnoteFile{Ma:2011:WSDM}

\bibitem{Yang:2011:WWW}
Yang SH, Long B, Smola A, Sadagopan N, Zheng ZH, et~al. (2011) Like like alike:
  joint friendship and interest propagation in social networks.
\newblock In: Proceedings of the 20th international conference on World Wide
  Web. WWW '11, pp. 537--546.
\bibAnnoteFile{Yang:2011:WWW}

\bibitem{Yuan:2011:RS}
Yuan Q, Chen L, Zhao SW (2011) Factorization vs. regularization: fusing
  heterogeneous social relationships in top-n recommendation.
\newblock In: Proceedings of the fifth ACM conference on Recommender systems.
  RecSys '11, pp. 245--252.
\bibAnnoteFile{Yuan:2011:RS}

\bibitem{Zhou:Plos:2011}
Zhou T, Medo M, Cimini G, Zhang ZK, Zhang YC (2011) Emergence of scale-free
  leadership structure in social recommender systems.
\newblock PLoS ONE 6: e20648.
\bibAnnoteFile{Zhou:Plos:2011}

\bibitem{Shang:EPL:2010}
Shang MS, L\"u LY, Zhang YC, Zhou T (2010) Empirical analysis of web-based
  user-object bipartite networks.
\newblock EPL 90: 48006.
\bibAnnoteFile{Shang:EPL:2010}

\bibitem{Zhou:NJP:2009}
Zhou T, Su RQ, Liu RR, Jiang LL, Wang BH, et~al. (2009) Accurate and diverse
  recommendations via eliminating redundant correlations.
\newblock New Journal of Physics 11: 123008.
\bibAnnoteFile{Zhou:NJP:2009}

\end{thebibliography}

\newpage

\section*{Figure Legends}

\section*{Tables}
\begin{table}[!ht]
\centering
\caption{The statistics of Amazon and Bookcross datasets. The sparsity is obtained by $\frac{\#links}{N\times M}$, where $N$ and $M$ are the number of users and items, respcetively.}\label{Tab:data_statistics}
\begin{tabular}{l|c|c|c|c}
\hline \hline
     Dataset & \#user & \#objects & \#links & sparsity \\[2pt]
    \hline
    Amazon & 50000 & 54,152 & 283,382 & $1.05\times 10^{-4}$  \\[2pt]
    Bookcross & 21122 & 203,373 & 504,643 & $1.17\times 10^{-4}$ \\[2pt]
    \hline\hline
\end{tabular}
\end{table}

\begin{table*}[!ht]
\centering
\caption{The optimal parameter defined in algorithms for \emph{Recall} and \emph{Ranking score}.}\label{Tab:optimal_parameter}
\begin{tabular}{cccccc}
\hline \hline
    \multicolumn{6}{c}{Amazon}  \\
    \hline
    & & SLD-T & RENBI  & U-SLD & O-SLD\\[2pt]
    \hline
    \multirow{2}{*}{Recall} & T & 2 & - & - & -\\
                    & $\theta$ & - & 2 & -0.2 & -0.3 \\[2pt]
    \hline
    \multirow{2}{*}{Ranking socre} & T & 5 & - & - & - \\
                                   & $\theta$ & - & 2 & -1.4 & -0.5 \\[2pt]
    \hline \hline
    \multicolumn{6}{c}{Bookcross}  \\[2pt]
    \hline
    & & SLD-T & RENBI & U-SLD & O-SLD- \\[2pt]
    \hline
    \multirow{2}{*}{Recall} & T & 2 & - & - & - \\
                    & $\theta$ & - & 1.0 & -0.5 & -0.2 \\[2pt]
    \hline
    \multirow{2}{*}{Ranking socre} & T & 5 & - & - & - \\
                                   & $\theta$ & - & 2 & -1.5 & -0.7 \\ [2pt]
    \hline\hline
\end{tabular}
\end{table*}

\end{document}